\documentclass[english,aps,two column]{revtex4-1}
\usepackage{ae,aecompl}
\usepackage{beramono}
\usepackage[T1]{fontenc}
\usepackage[latin9]{inputenc}
\usepackage{amssymb}
\usepackage{graphicx}
\usepackage{esint}
\usepackage{babel}
\begin{document}

\title{Signature of pseudo-diffusive transport in mesoscopic topological
insulators}

\author{Saurav Islam$^{1}$, Semonti Bhattacharyya$^{1}$, Hariharan Nhalil$^{1}$,
Suja Elizabeth$^{1}$, Arindam Ghosh$^{1,2}$}

\affiliation{$^{1}$Department of Physics, Indian Institute of Science, Bangalore: 560012}

\affiliation{$^{2}$Center for Nanoscience and Engineering, Indian Institute of
Science, Bangalore: 560012}
\email{isaurav@iisc.ac.in}

\begin{abstract}
One of the unique features of Dirac Fermions is pseudo-diffusive transport
by evanescent modes at low Fermi energies when disorder is low. At
higher Fermi energies \textit{i.e.} carrier densities, the electrical
transport is diffusive in nature and the propagation occurs via plane-waves.
In this study, we report the detection of such evanescent modes in
the surface states of topological insulator through $1/f$ noise for
the first time. While signatures of pseudo-diffusive transport have
been seen experimentally in graphene, such behavior is yet to be observed
explicitly in any other system with a Dirac dispersion. To probe this,
we have studied $1/f$ noise in topological insulators as a function
of gate-voltage, and temperature. Our results show a non-monotonic
behavior in $1/f$ noise as the Fermi energy is varied, suggesting
a crossover from pseudo-diffusive to diffusive transport regime in
mesoscopic topological insulators. The temperature dependence of noise
points towards conductance fluctuations from quantum interference
as the dominant source of the noise in these samples.
\end{abstract}
\maketitle
Topological insulators\ (TIs), with their spin-polarized, topologically
protected, linear, metallic surface states, act as the perfect playground
for investigating a plethora of fundamental phenomena\ \cite{hasan2010colloquium,hsieh2008topological,roushan2009topological,konig2007quantum,zhang2009topological}.
These surface carriers obey the Dirac equation for massless Fermions,
where the Hamiltonian of the system is given by $H=\hbar v_{F}\overrightarrow{\sigma}\cdot\overrightarrow{k}$.
Here $v_{F}$, $\vec{\sigma}$, and $\vec{k}$ refer to the Fermi
velocity, spin matrices, and momentum respectively. Due to the massless
nature of the charge carriers, the screening properties of Dirac materials
such as TIs or graphene, are also significantly different from other
traditional 2D electron systems, and the potential due to charged
disorder remains long-ranged even after screening is taken into account
in Dirac materials\ \cite{rossi2012universal}. Another key feature
of these materials is that it is possible to reach $\langle n\rangle=0$,
without opening up a band-gap, even though strong carrier inhomogenities
in the form of electron-hole puddles might be present around charge
neutrality point or the Dirac point\ \cite{beidenkopf2011spatial}.
The electrical transport properties of these classes of materials
near the Dirac point, where the Fermi surface diminishes to a point,
has been a matter of intense discourse, and has led to several fascinating
discoveries in the context of graphene, such as dissipative quantum
Hall effect, minimum conductivity, and pseudo-diffusive transport\ \cite{pal2011microscopic,geim2004,novoselov2005two,neto2009electronic,tworzydlo2006sub,katsnelson2006zitterbewegung,cuevas2006subharmonic,akhmerov2007pseudodiffusive,titov2006josephson,dicarlo2008shot,danneau2008shot,miao2007phase,heersche2007bipolar,du2008josephson,kumaravadivel2016signatures,abanin2007dissipative,abanin2011giant,peres2010colloquium,tan2007measurement,checkelsky2009divergent,jiang2007quantum,feldman2009broken,du2009fractional}.
Accessing the Dirac point in TIs compared to graphene has been a challenge
due to high doping from bulk defects as well as the substrate, thus
making it difficult to probe the intriguing properties of Dirac Fermions
in TIs including the origin of $1/f$ noise. Previous investigation
of $1/f$ noise in TIs have revealed the role of bulk disorder-mediated
Hooge type mobility fluctuation type noise in $100$\ nm thick mesoscopic
samples and correlated mobility-number density fluctuation model to
be the dominant mechanism in large area epitaxially grown samples\ \cite{bhattacharyya2015bulk,bhattacharyya2016resistance,islam2017bulk,islam2018universal,ghatak2011nature,pal2011microscopic,ghatak2014microscopic,shamim2016ultralow}.
However, the origin of $1/f$ noise in TIs in thin (thickness $d\sim10$\ nm)
mesoscopic samples, especially near the Dirac point, also remains
a matter of debate. In this manuscript, we have explored the origin
of $1/f$ noise in mesoscopic samples, where we have access to the
Dirac point also. Our investigation has revealed a non-monotonic dependence
of $1/f$ noise magnitude on the carrier number density, which is
a strong function of temperature as well, suggesting a crossover from
pseudo-diffusive to diffusive transport, and the conductance fluctuations
from quantum interference effects as the main source of noise in these
types devices at low $T$.

\begin{figure}
\includegraphics[width=1\columnwidth]{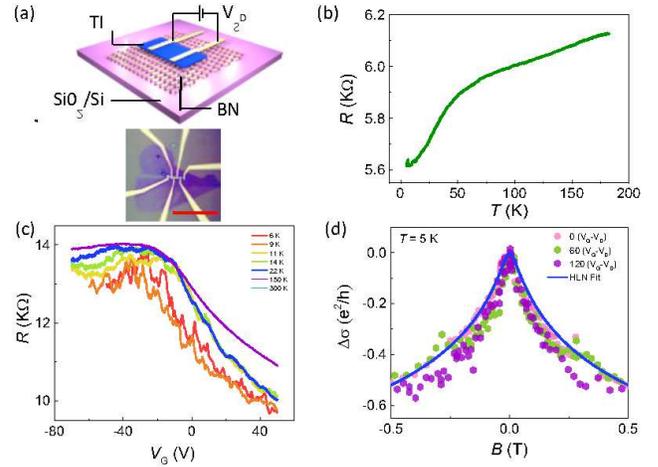}\caption{\textbf{Device characteristics} (a) Schematic of a typical TI-FET
(top). Optical micrograph of an actual device (below). (b) Resistance
vs temperature of sample BT$1$, showing a weak dependence on $T$,
indicating metallic behavior and the dominance of surface states in
the transport. (c) Resistance vs gate-voltage of sample BT$1$, at
different $T$, showing ambipolar transport. (d) Weak-antilocalization
in TIs, characterized by a cusp in the correction to conductivity
at $B=0$\ T. The solid lines are fits to the data using Eq\ \ref{eq:HLN}.\label{fig:Device-characteristics}}
\end{figure}

\begin{figure}
\includegraphics[width=1\columnwidth]{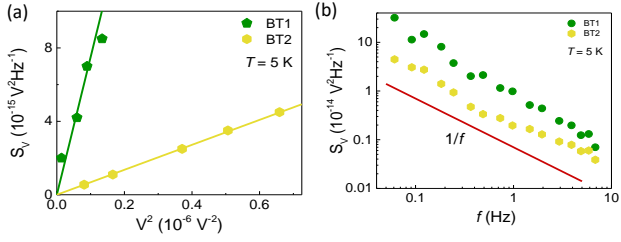}\caption{\textbf{Noise measurements:} (a) Schematic of the experimental setup
for measuring $1/f$ noise. (b) S$_{V}$ vs $V^{2}$ at $T=5$\ K
for both samples BT$1$ and BT$2$, showing a quadratic behavior,
implying that the response is in the linear regime. (c) S$_{V}$ as
a function of frequency, showing $1/f$ behaviour. \label{fig:noise}}
\end{figure}
The devices studied in this paper were fabricated using the TI Bi$_{1.6}$Sb$_{0.4}$Te$_{2}$Se,
which was exfoliated from a single crystal onto a SiO$_{2}$/Si wafer
using Scotch tape technique\ \cite{taskin2011observation,geim2004,novoselov2005two}.
Due to compensation doping, the quarternary alloy Bi$_{1.6}$Sb$_{0.4}$Te$_{2}$Se
has an insulating bulk, resulting in enhanced surface transport\ \cite{taskin2011observation,ren2011optimizing,bhattacharyya2015bulk}.
Here below temperature $T=50$\ K for samples with thickness, $d\leq100$\ nm,
the current is essentially carried by the surface carriers\ \cite{bhattacharyya2015bulk}.
The atomically flat boron nitride\ (BN) substrate\ (Fig.\ \ref{fig:Device-characteristics}a),
significantly reduces the effect of dangling bonds and charged traps
of the SiO$_{2}$ substrate on the electrical transport in the TI
channel\ \cite{dean2010boron,karnatak2016current,islam2018universal}.
The hetero-structure was then finally assembled onto a pre-patterned
heavily hoped SiO$_{2}$/Si substrate, with the $285$\ nm thick
SiO$_{2}$ acting as a back gate dielectric, using a home-made transfer
technique. The sample contacts were patterned by standard electron-beam
lithography, followed by thermal evaporation of $5/40$\ nm Cr/Au
(Fig.\ \ref{fig:Device-characteristics}a). A layer of the polymer
PMMA\ (poly(methylmethacrylate)) was coated on the samples, which
ensured that the surface integrity is preserved throughout the measurement
cycle. The measurements reported in this manuscript were performed
on two identically prepared samples, BT$1$ and BT$2$, in a home-built
variable temperature cryostat. The resistivity measurements were performed
using a low frequency AC-four probe technique with carrier frequency
of $18$\ Hz with an excitation current of $100$\ nA. 

The resistance\ ($R$) vs temperature\ ($T$) shows metallic behavior,
implying the predominance of surface states in the transport, as expected
for $10$\ nm thin TIs flakes\ (Fig.\ \ref{fig:Device-characteristics}b)\ \cite{bhattacharyya2015bulk}.
Fig.\ \ref{fig:Device-characteristics}c shows the $R$ vs $V_{G}$,
where a maximum in the resistance at $V_{G}\approx-40$\ V at $T=5$\ K
represents the Dirac point. The asymmetry in the $R$-$V_{G}$ on
the electron and holes sides may arise due to asymmetry in the band-structure
itself\ \cite{adam2012two}. The typical mobility extracted from
the $R-V_{G}$ graph is $\sim100$\ cm$^{2}$V$^{-1}$s$^{-1}$.
Fig.\ \ref{fig:Device-characteristics}d shows magneto-resistance\ (MR)
behavior of BT$1$ at $V_{G}-V_{D}=0$\ V, $60$\ V and $120$\ V,
characterized by a cusp in the quantum correction to conductivity
$\triangle\sigma$ at $B=0$\ T\ \cite{zhang2013weak,he2011impurity,kim2011thickness,bansal2012thickness}.
This demonstrates weak-antilocalization phenomenon, as expected for
spin-momentum locked TI surface states, resulting from an additional
$\pi$ Berry phase between the back-scattered, time reversed path
of the carriers leading to negative magneto-conductance. The magneto-conductance
data can be fitted with the Hikami-Larkin-Nagaoka (HLN)\ \cite{hikami1980spin,bao2012weak}
equation for diffusive metals with high spin orbit coupling\ $(\tau_{\phi}>>\tau_{so},\tau_{e})$:

\begin{equation}
\triangle\sigma=-\alpha\frac{e^{2}}{\pi h}\left[\psi\left(\frac{1}{2}+\frac{B_{\varphi}}{B}\right)-\ln\left(\frac{B_{\varphi}}{B}\right)\right]\label{eq:HLN}
\end{equation}
where $\tau_{\phi}$, $\tau_{so}$, $\tau_{e}$ are the phase coherence
or dephasing time, spin-orbit scattering time and elastic scattering
time respectively, $\psi$ is the digamma function and $B_{\phi}$
is the phase breaking field. Here $\alpha$ and $B_{\phi}$ are fitting
parameters. The phase coherence length $l_{\phi}^{MR}$ can be extracted
using $l_{\phi}^{MR}=\sqrt{\hbar/4eB_{\phi}}$. The $l_{\phi}$ obtained
from the fit was $\sim180$\ nm at $T=5$\ K for $V_{G}-V_{D}=0$\ V. 

\begin{figure*}
\includegraphics[width=1\textwidth]{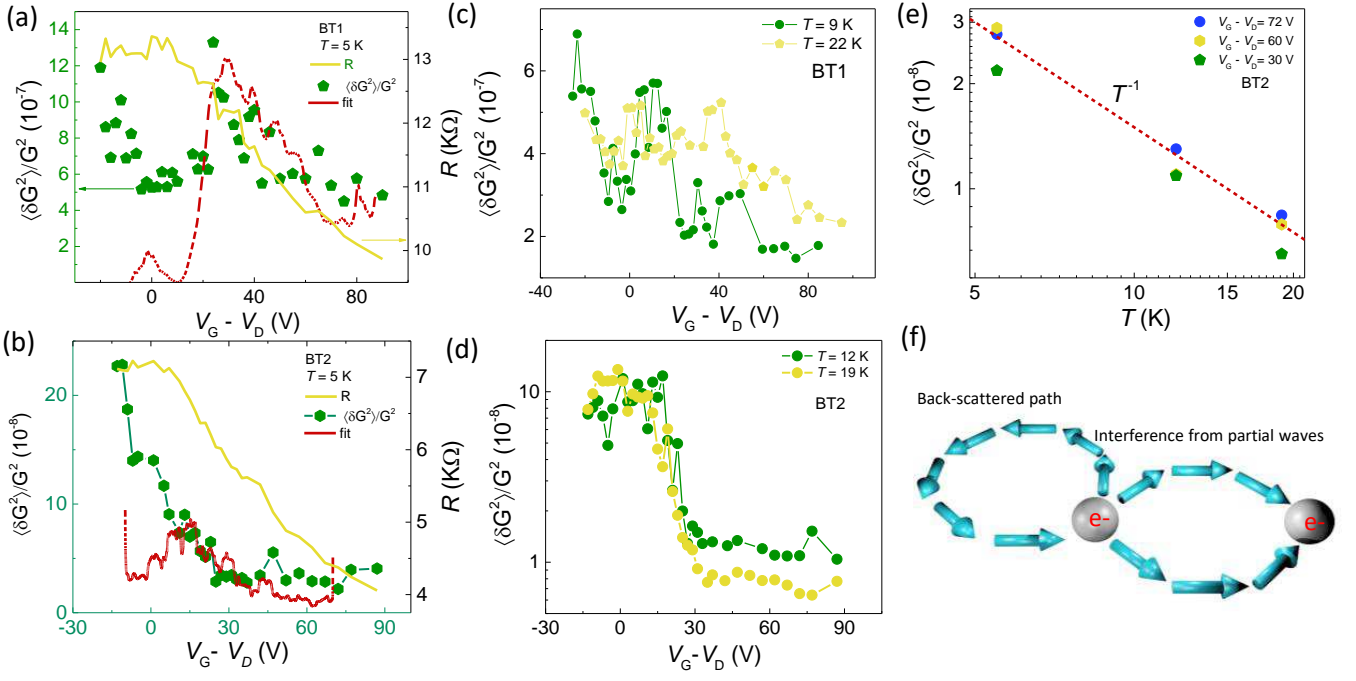}\caption{\textbf{1/$f$ noise measurements:} (a) Integrated noise magnitude
$\frac{\langle\delta G^{2}\rangle}{G^{2}}$ as a function of $V_{G}$
for sample BT$1$. The dashed red line shows fit according to Eq.\ \ref{eq:jayaraman}.
(b) $\frac{\langle\delta G^{2}\rangle}{G^{2}}$ as a function of $V_{G}$
for sample BT$2$. The dashed red line shows fit according to Eq.\ \ref{eq:jayaraman}
for sample BT$2$. (c)$V_{G}$-dependence of $\frac{\langle\delta G^{2}\rangle}{G^{2}}$
at $T=9$\ K and $22$\ K for sample BT$1$. (d) $\frac{\langle\delta G^{2}\rangle}{G^{2}}$
as a function of $V_{G}$ for $T=12$\ K and $19$\ K for sample
BT$2$. (e) $\frac{\langle\delta G^{2}\rangle}{G^{2}}$ vs $T$ for
sample BT$2$, at $V_{G}-V_{D}=30$~V,$60$\ V, and $70$\ V, showing
$1/T$ dependence, indicating the origin of $1/f$ noise to be from
universal conductance fluctuations. (f) Schematic illustration of
universal conductance fluctuations, which originated from quantum
interference effect.\label{fig:1/-noise-measurements:}}
\end{figure*}
To extract the magnitude of $1/f$ noise of the samples accurately,
we have utilized a AC four-probe Wheatstone bridge technique\ \cite{scofield1987ac,weissman1996low,dutta1981low}.
The voltage fluctuations were recorded as a function of time using
a 16-bit digitizer. This was followed by digital processing of the
time-series data to obtain the power spectral density (PSD, $S_{V}$)
as a function of frequency ($f$) (Fig\ \ref{fig:noise}a). In both
the devices BT1 and BT2, $S_{V}\propto1/f^{\alpha}$, and the exponent
of the frequency, $\alpha\approx1$. $S_{V}$ depends on the the bias
($V$) in a quadratic manner, which ensured that all the measurements
were performed in the Ohmic regime\  (Fig\ \ref{fig:noise}b). 

The $V_{G}$-dependence of the integrated noise magnitude ($\frac{\langle\delta G^{2}\rangle}{G^{2}}=\int\frac{S_{v}}{V^{2}}df$)
at $T=5$\ K, shown in Fig.\ \ref{fig:1/-noise-measurements:}a
and Fig.\ \ref{fig:1/-noise-measurements:}b for samples BT$1$ and
BT$2$. Although these two samples were identically prepared from
the same bulk crystal, and show similar average electrical characteristics\ \cite{islam2018universal},
they demonstrate contrasting behaviors in the $V_{G}$ dependence
of noise. Whereas $\frac{\langle\delta G^{2}\rangle}{G^{2}}$ vs $V_{G}$
for sample BT$1$ displays a M-shaped curve with a dip around the
Dirac point\ ($\mid V_{G}-V_{D}\mid=20$\ V)\ (Fig.\ \ref{fig:1/-noise-measurements:}a),
the identically prepared device BT$2$ shows a monotonic reduction
as $V_{G}$ is tuned away from the Dirac point, as demonstrated in
Fig.\ \ref{fig:noise}b. The non-monotonic behavior of $1/f$ noise
previously reported in the context of graphene\ \cite{pal2011microscopic}
and in TIs\ \cite{bhattacharyya2015bulk}, has been attributed to
the interplay of charge exchange noise (originating due to exchange
of carriers between the channel and the surrounding environment) and
configuration noise (arising due to potential fluctuations due to
reorganization of trapped charges). Incase of graphene, however, this
dip in noise across the Dirac point persists till room temperature,
while for mesoscopic TI-FETs, this is a very strong function of $T$,
and persists only till $T=14$\ K in sample BT$1$. We have fitted
the $V_{G}$-dependence of the normalized noise magnitude data using
the framework of correlated mobility-number density fluctuations model\ \cite{Jayaraman1989,islam2017bulk},
which is known to be the dominant mechanism of noise in large-area,
thin ($\sim10$\ nm) TIs, where the effect of conductance fluctuations
are suppressed to a large $L/l_{\phi}$ ratio. The total noise can
be expressed as, 
\begin{equation}
\frac{S_{V}}{V^{2}}=\frac{D_{it}k_{B}T}{dWL}\left(\frac{d\sigma}{dn}\right)^{2}\left(\frac{J_{1}}{\sigma^{2}}+\frac{J_{2}}{\sigma}+J_{3}\right)\label{eq:jayaraman}
\end{equation}
where $J_{1}=\frac{1}{8\alpha}$ represents a pure number fluctuation,
$J_{3}=\int A^{2}(x)\frac{\tau_{_{T}}}{1+\left(2\pi f\tau_{_{T}}\right)^{2}}dx$
represents pure mobility fluctuations and $J_{2}=\int2A(x)\frac{\tau_{_{T}}}{1+\left(2\pi f\tau_{_{T}}\right)^{2}}dx$
represents combined number and mobility fluctuations ($\alpha$ is
the decay constant for the spatially decaying time constant $\tau_{_{T}}$
of a typical trapping event and $A(x)$ is the scattering constant)
and can be evaluated using phenomenological values\ \cite{Jayaraman1989}.
$D_{it}$, $k_{B}$, $W$, $L$, $\sigma$, $n$, $x$ are the areal
trapped charge density per unit energy, Boltzmann constant, width
of the channel, length of the channel, conductance and number density
of charge carriers, axis in the direction perpendicular to the channel
respectively, $f=1$\ Hz frequency and $d=1$\ nm is the distance
over which the tunneling is effective. As is evident from the fit,
this framework does not satisfactorily explain the observed nature
of $1/f$ noise in mesoscopic samples, implying that the dominant
source of $1/f$ in mesoscopic samples and large area TI samples are
different\ (Fig.\ \ref{fig:1/-noise-measurements:}a-b). Such behavior
of $1/f$ noise on the number density have been predicted theoretically
for Dirac fermions for long-range as well as Gaussian disorder, due
to a crossover from pseudo-diffusive to diffusive transition, which
we believe is the scenario here\ \cite{rossi2012universal}. In the
pseudo-diffusive regime, the transport in the channel occurs through
quantum tunneling of evanescent modes. However, due to the presence
of disorder, the system is driven into a diffusive metal phase, with
the propagation occurring via plane waves. Although signatures of
pseudo-diffusive transport has been reported in graphene\ \cite{pal2011microscopic,geim2004,novoselov2005two,neto2009electronic,tworzydlo2006sub,katsnelson2006zitterbewegung,cuevas2006subharmonic,akhmerov2007pseudodiffusive,titov2006josephson,dicarlo2008shot,danneau2008shot,miao2007phase,heersche2007bipolar,du2008josephson,kumaravadivel2016signatures,dufouleur2017weakly},
there is no such clear signature in TIs. In the pseudo-diffusive regime,$\langle\delta G^{2}\rangle$
enhances rapidly in magnitude compared to $\langle G\rangle$ with
increasing $n$, while in the diffusive regime, $\langle\delta G^{2}\rangle$
is almost constant whereas $\langle G\rangle$ increases. This leads
to a non-monotonic dependence of $1/f$ noise magnitude on $n$, which
is a generic property of crossover between these two regimes. 

To gain further insight into the origin of $1/f$ noise in mesoscopic
TI-FETs, we have performed $V_{G}$-dependence of noise at various
temperatures for both samples. The non-monotonic behavior of $1/f$
noise in sample BT$1$ shows a strong $T$-dependence with the peak
almost disappearing for $T>20$\ K\ (Fig.\ \ref{fig:1/-noise-measurements:}c).
The $V_{G}$-dependence of noise in sample BT$2$ shows a monotonic
decrease with number density at all temperatures. The $T$-dependence
of noise for sample BT$2$ at various gate-voltages is shown in Fig.\ \ref{fig:1/-noise-measurements:}e.
The magnitude of noise, reduces as the $T$ is increased (Fig.\ \ref{fig:1/-noise-measurements:}e),
contrary to what has been observed in MBE grown TIs before, where
the noise magnitude increases due to scattering from thermally activated
defects\ \cite{islam2017bulk}. The noise magnitude, as shown in
Fig.\ \ref{fig:1/-noise-measurements:}e, for BT$2$, reduces as
$\sim T^{-1}$. Such a $T$-dependence of noise can be explained using
the framework of universal conductance fluctuations. For $T\rightarrow0$,
UCF magnitude $\left\langle \delta G^{2}\right\rangle ^{\frac{1}{2}}\rightarrow e^{2}/h$,
while at finite temperature $\langle\delta G^{2}\rangle\simeq\left(\frac{e^{2}}{h}\right)^{2}\alpha(k_{F}\delta r)\frac{1}{k_{F}l}\frac{L_{y}}{L_{x}^{3}}n_{s}(T)l_{\phi}^{4}$,
where $k_{F}$, $l$, $L_{x}$ and $L_{y}$ are the Fermi wave-vector,
mean free path and sample dimensions in $x$ and $y$ directions respectively\ \cite{birge1990conductance,feng1986sensitivity,shamim2017dephasing}.
$\alpha(x)$ represents the change of the phase of electron wave-function
due to scattering by a moving impurity at a distance $\delta r$,
and $n_{s}(T)$ is the number of active scatteres. For electron-electron
interaction mediated dephasing, $l_{\phi}^{2}\propto1/T$ and $n_{s}(T)\propto T$,
we have $\langle\delta G^{2}\rangle\propto l_{\phi}^{4}n_{s}(T)\propto1/T$,
as observed\ \cite{birge1990conductance,feng1986sensitivity,shamim2017dephasing,altshuler1982effects}.
While the overall noise magnitude for sample BT$1$ reduces at the
$T$ is increased, there is no specific trend which is observed, and
the noise in the data prevents a conclusive claim in this particular
sample.

\begin{figure}
\includegraphics[width=1\columnwidth]{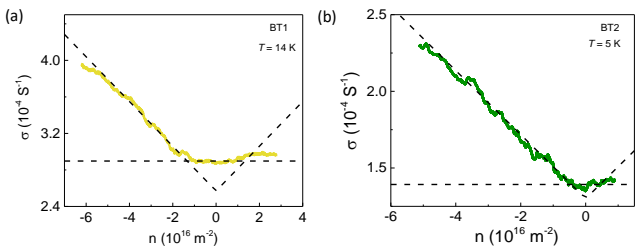}\caption{\textbf{Impurity density calculation: }(a) (a) Conductance ($\sigma=\frac{L}{RW}$)
vs. calculated number density ($n_{calc}=\frac{C\left(V_{G}-V_{D}\right)}{e}$)
at $T=6$\ K for sample BT$1$. The black lines are fit of this data
according to the Eq.\ \ref{eq:coulomb imp 1} and \ref{eq:coulomb imp 2}.
(b) $\sigma$ vs $n$ for sample BT$2$. The solid lines are fits
to the data according to Eq.\ \ref{eq:coulomb imp 1} and \ref{eq:coulomb imp 2}.
\label{fig:Impurity-density-calculation} }
\end{figure}

Taking into consideration these results, we believe that the origin
of $1/f$ noise in thin, mesoscopic samples of TIs can be attributed
to universal conductance fluctuations, which arises due to quantum
interference effects\ \cite{lee1985universal,lee1987universal,feng1986sensitivity,islam2018universal,islam2017bulk,bhattacharyya2016resistance}
and is schematically shown in Fig.\ \ref{fig:1/-noise-measurements:}f.
The charge carriers undergo multiple elastic scatterings from impurities,
defects or boundaries, and follow trajectories which are a strong
function of disorder configuration, Fermi energy, and magnetic field.
Interference between these trajectories, which can involve back-scattered
carriers or interference between partial waves between two points
having different paths leads to conductance fluctuations, whose noise
spectra is $1/f$ in nature\ \cite{lee1985universal}. These conductance
fluctuations are the dominant source of $1/f$ noise in mesoscopic
topological insulators at low $T$. 

To verify whether this is the dominant mechanism, we have fitted $\sigma$-$n$
data (Fig\ \ref{fig:Impurity-density-calculation}a), where $\sigma=\frac{L}{RW}$
and $n=\frac{C_{S}\left(V_{G}-V_{D}\right)}{e}$, within the framework
of charge-impurity limited scattering of Dirac fermions\ \cite{kim2012surface},
where

\begin{eqnarray}
\sigma\sim E\left|\frac{n}{n_{imp}}\right|\left[e^{2}/h\right]\ \  & \mathrm{for}\ \  & n>n^{*}\label{eq:coulomb imp 1}
\end{eqnarray}

and

\begin{eqnarray}
\sigma\sim E\left|\frac{n^{*}}{n_{imp}}\right|\left[e^{2}/h\right]\ \  & \mathrm{for}\ \  & n<n^{*}\label{eq:coulomb imp 2}
\end{eqnarray}
where $n^{*}$ is the residual carrier density in electron and hole
puddles, and $E$ is a constant depending on the Wigner\textendash Seitz
radius $r_{s}$. The extracted value of number density of Coulomb
traps in sample BT$1$ is $n_{imp}=1.5\times10^{16}$\ m$^{-2}$,
while for BT$2$, $n_{imp}=5\times10^{16}$\ m$^{2}$, which matches
well with the theoretically predicted values. The density of electron-hole
puddles is $n^{*}=7\times10^{14}$\ m$^{-2}$ and $n^{*}=5\times10^{15}$\ m$^{-2}$
for samples BT$1$ and BT$2$ respectively. This difference in impurity
density is reflected in the the qualitative nature of $V_{G}$-dependence
of noise as seen in Fig.\ \ref{fig:1/-noise-measurements:}a-b, thereby
providing further support to the observation of pseudo-diffusive transport
in device BT$1$.

In summary, we have measured time-dependent voltage fluctuations to
extract the magnitude of $1/f$ noise in mesoscopic topological insulators
devices as a function of gate-voltage and temperature. The temperature
dependence implies that the noise originates from universal conductance
fluctuations due to quantum interference effects. More importantly,
the non-monotonic dependence of noise on the number density in the
low disordered samples signifies a crossover from pseudo-dissusive
to diffusive transport regime, a phenomena unique to Dirac Fermions.
\bibliographystyle{apsrev4-1}

\end{document}